# Electrical Control of Neutral and Charged Excitons in a Monolayer Semiconductor

Jason S. Ross[1*], Sanfeng Wu[2*], Hongyi Yu[3], Nirmal J. Ghimire[4,5], Aaron M. Jones[2], Grant Aivazian[2], Jiaqiang Yan[5,6], David G. Mandrus[4,5,6], Di Xiao[7], Wang Yao[3], Xiaodong Xu[1,2†]

[1] Department of Material Science and Engineering, University of Washington, Seattle, Washington 98195, USA

[2] Department of Physics, University of Washington, Seattle, Washington 98195, USA

[3] Department of Physics and Center of Theoretical and Computational Physics, The University of Hong Kong, Hong Kong, China

[4] Department of Physics and Astronomy, University of Tennessee, Knoxville, Tennessee 37996, USA

[5] Materials Science and Technology Division, Oak Ridge National Laboratory, Oak Ridge, Tennessee, 37831, USA

[6] Department of Materials Science and Engineering, University of Tennessee, Knoxville, Tennessee, 37996, USA

[7] Department of Physics, Carnegie Mellon University, Pittsburg, PA 15213, USA

* These authors contributed equally to the work.

† Email: xuxd@uw.edu

**Abstract: Monolayer group VI transition metal dichalcogenides have recently emerged as semiconducting alternatives to graphene in which the true two-dimensionality (2D) is expected to illuminate new semiconducting physics. Here we investigate excitons and trions (their singly charged counterparts) which have thus far been challenging to generate and control in the ultimate 2D limit. Utilizing high quality monolayer molybdenum diselenide ($MoSe_2$), we report the unambiguous observation and electrostatic tunability of charging effects in positively charged ($X^+$), neutral ($X^o$), and negatively charged ($X^-$) excitons in field effect transistors via photoluminescence. The trion charging energy is large (30 meV), enhanced by strong confinement and heavy effective masses, while the linewidth is narrow (5 meV) at temperatures below 55 K. This is greater spectral contrast than in any known quasi-2D system. We also find the charging energies for $X^+$ and $X^-$ to be nearly identical implying the same effective mass for electrons and holes.**

Introduction

Above bandgap photo-excitation creates electrons and holes in the conduction and valence bands respectively. If the screening is weak enough the attractive Coulomb interaction between one electron and one hole creates a bound quasi-particle known as a neutral exciton ($X^o$) which has an energy structure similar to a neutral hydrogen atom. Excitons can further become charged by binding an additional electron ($X^-$) or hole ($X^+$) to form charged 3-body excitons analogous to $H^-$ or $H_2^+$ respectively[1–3]. These exciton species are elementary quasi-particles describing the electronic response to optical excitation in solids and are integral to many optoelectronic applications from solar cells and light emitting diodes (LEDs)[4] to optical interconnects[5] and quantum logical devices[6,7].

The main arena for the exploration of excitonic physics has been 3D semiconductors and their heterostructures that form quasi-2D quantum wells where the carrier wavefunction typically occupies a few tens to thousands of atomic layers. Observation and control of excitons in truly 2D systems has been a long pursued goal, largely motivated by the enhancement in exciton and trion binding energies in the strict 2D limit[8]. In addition, unlike semiconductor heterostructures, the close proximity of external stimuli to the exciton wavefunction can offer unprecedented tunability and diversifies the applications possible in devices made with 2D semiconductors.

Here we report the experimental observation and control of the fascinating excitonic physics in a 2D semiconductor by utilizing high quality monolayer molybdenum diselenide ($MoSe_2$). $MoSe_2$ belongs to the group-VI transition metal dichalcogenides which form in layers weakly bound to each other by Van der Waals forces. The monolayers have received much attention recently as they make up a new class of 2D semiconductors with a direct bandgap in the visible frequency range and are predicted to exhibit coupled spin-valley physics[9]. Recent progresses focus on $MoS_2$, including the demonstration of having a direct bandgap[10,11], high mobility electronics[12], optical generation of valley polarization[13–15], and electrical control of Berry phase properties[16].

Using a back-gated field effect transistor (FET) device, we demonstrate the reversible electrostatic tunability of the exciton charging effects from positive ($X^+$) to neutral ($X^o$) and to negative ($X^-$). We observe a large trion binding energy of 30 meV with a narrow emission linewidth of 5 meV. These narrow, well separated features have temperature dependence of typical 2D excitons and exist at high temperature suggesting remarkable stability. Interestingly, the binding energies of $X^+$ and $X^-$ are similar implying that low energy electrons and holes in $MoSe_2$ have the same effective mass. Our work demonstrates that monolayer $MoSe_2$ is a true 2D semiconductor opening the door for the investigation of phenomena such as exciton condensation[17–19] and the Fermi-edge singularity[20,21], as well as for a new generation of optoelectronic devices such as LEDs and excitonic circuits[5].

## Results

**Crystal structure and spectral features of monolayer MoSe$_2$.** In monolayer MoSe$_2$, Mo and Se atoms form a 2D hexagonal lattice with trigonal prismatic coordination (Fig. 1a). First principles calculations show that it has a direct bandgap at the corners (K points) of the first Brillouin zone (Fig. 1b). The curvature of the bands suggests comparable effective mass for low energy electrons and holes at K points (Fig. 1c) [9]. These band edge electrons and holes near K points are predominantly from the *d*-orbitals of Mo atoms. Their wavefunctions are calculated to be strongly confined in the Mo layer within a length scale of ~0.2 nm in the out-of-plane direction. Monolayer MoSe$_2$ is thus an ideal nanomaterial for exploring excitonic physics in the ultimate 2D limit.

We use mechanical exfoliation to obtain monolayer MoSe$_2$ on 300 nm SiO$_2$ on n+ doped Si and atomic force microscopy (AFM) to identify the layer thickness[22]. Figure 1d,e show the optical micrograph and the corresponding AFM image of a representative sample, in which the monolayer thickness of ~0.7 nm is identified[12] (Fig. 1f). Standard electron beam lithography (EBL) is used to fabricate monolayer field effect transistors (Fig. 1g). With the contacts simply grounded, the n+ Si functions as a back gate providing uniform electrostatic doping in the MoSe$_2$ (Fig. 1h).

The excitonic features of MoSe$_2$ are investigated by differential reflectance and micro-photoluminescence (µ-PL) measurements (Methods). Figure 2 shows the results from an unpatterned MoSe$_2$ sample, S1. At 20 K, we observe two main features associated with the A and B excitons in the differential reflection spectrum[10,11,14,23–25] (Fig. 2a). The presence of A and B excitons has been attributed to spin-orbit coupling induced valence band splitting in bulk[24]. The observed energy difference of ~200 meV agrees well with the calculated splitting (180 meV) in monolayers (Fig. 1c).

With the same sample and temperature under 2.33 eV laser excitation, the PL spectrum does not show a measurable feature which can be attributed to the B exciton, likely since it is not the lowest energy transition. Instead, we observe two pronounced peaks at 1.659 and 1.627 eV in the vicinity of the A exciton (Fig. 2b). Note that the PL spectrum lacks the broad low energy peak observed in MoS$_2$ which has been attributed to defect-related, trapped exciton states[10,11,14,25].

The striking spectral features demonstrate the high quality of our MoSe$_2$ samples (Methods) and provide strong evidence for monolayer MoSe$_2$ being a direct bandgap semiconductor (Supplementary Figure S1) where the two distinct transitions are excitons. The higher energy emission at 1.659 eV is the neutral exciton, X$^o$, and the lower energy peak is a trion[3]. In unpatterned samples, we assume the trion to be X$^-$ since all measured devices show *n*-doped characteristics (Supplementary Figure S2). All measured unpatterned samples show a binding energy, which is the energy difference between trion and X$^o$, of ~30 meV (Fig. 2b inset). This is more than twice typical numbers reported in GaAs quantum wells[3,26,27] and similar to a recent mention in MoS$_2$[14].

**Gate dependence of MoSe$_2$ photoluminescence.** In order to confirm the above assignment and control the exciton charging effects, we performed gate dependent PL measurements using monolayer MoSe$_2$ FETs. Here, the excitation laser is at 1.73 eV for better resonance with the luminescent states. Figure 3a shows a color map of the PL spectrum of device D1 at 30 K as a function of back-gate voltage, V$_g$, in which we clearly observe four spectral features whose intensities strongly depend on V$_g$. Near zero V$_g$ the spectrum shows a broad low energy feature around 1.57 eV and a narrow high energy peak at 1.647 eV. With large V$_g$ of either sign, these peaks disappear and a single emission peak dominates the spectrum. Both peaks (at negative or positive V$_g$) have similar energies and intensities with the latter increasing with the magnitude of V$_g$.

This observed gate dependence confirms the assignment of states as labeled in Figure 3a. Since the broad low energy peak does not show up in unpatterned samples before FET fabrication, we attribute it to exciton states trapped to impurities (X$^I$)[10,11,14,25] which are likely introduced during EBL processing and are not the focus of this paper. The sharp peak at 1.647 eV is the X$^o$, slightly red-shifted compared to unpatterned samples. From the gate dependence, we identify the peaks near 1.627 eV as the X$^-$ and X$^+$ trions when V$_g$ is largely positive and negative respectively. Remarkably, these two distinct quasi-particles (X$^+$ and X$^-$) exhibit a nearly identical binding energy. The difference is within 1.5 meV over the whole applied *V$_g$* range. Since the binding energy of a trion is dependent on its effective mass, this observation implies that the electron and hole have approximately the same effective mass.

The gate dependent measurements unambiguously demonstrate the electrical control of exciton species in a truly 2D semiconductor, as illustrated in Figure 3b. The conversion from $X^o$ to trion can be represented as $e(h) + X^o \rightarrow X^- (X^+)$, where e and h represent an electron or a hole respectively. By setting $V_g$ to be negative, the sample is *p*-doped, favoring excitons to form lower energy bound complexes with free holes. As $V_g$ decreases, more holes are injected into the sample and all $X^o$ turn into $X^+$ to form a positively charged hole-trion gas. With positive $V_g$ a similar situation occurs with free electrons to form an electron-trion gas. In the following, we show that a standard mass action model can be used to describe the conversion dynamics.

Figure 3c shows the extracted $X^o$ (black) and trion (red) peak intensity as a function of $V_g$ where we have adjusted the negative $V_g$ data due to background signal. The plot shows that the maximum $X^o$ intensity is about equal to the saturated trion PL when $X^o$ vanishes. This observation indicates conservation of the total number of $X^o$ and trion in the applied voltage range and similar radiative decay rates for both quasi-particles. Thus the PL intensity represents the amount of the corresponding exciton species. Since the dynamic equilibrium of free electrons, holes, and excitons are governed by the rate equation and law of mass action[27], we calculate the gate dependent $X^o$ and trion abundance (Supplementary Figure S3), shown by the solid lines in Figure 3c, which agrees with the data.

In the simulation, we first fit the $X^o$ curve to obtain our two free parameters: the maximum background electron concentration $n_B^{max} = 3.6 \times 10^{10} \text{cm}^{-2}$ when the trion intensity saturates and the photo-excited electron concentration $n_P = 1.5 \times 10^{10} \text{cm}^{-2}$. We then fit the trion gate dependence with these parameters held fixed (Supplementary Note 1). The deviation in the trion experimental data from the calculated curve near zero $V_g$ is artificial due to the mutual background from $X^o$ and $X^I$. We note that this inferred electron concentration is much smaller than the product of the gate capacitance and $V_g$. This discrepancy can be attributed to the large contact resistance (Supplementary Figure S4) of the sample, which prevents the carrier concentration from reaching equilibrium on the experimental time scale at 30 K. We expect that future improved contact technologies will eliminate this effect.

**Temperature dependence of MoSe$_2$ photoluminescence.** The observed exciton states also show fine features consistent with 2D excitons, such as temperature dependent line shape, peak energy, and relative weight of $X^o$ and trion, which further supports the excitonic nature of this monolayer system (Fig. 4). Figure 4a shows the evolution of X⁻ and X⁰ (normalized PL) as a function of temperature in an unpatterned sample, S2, under 1.96 eV laser excitation. At low temperatures, we again observe a binding energy of 30 meV. As the temperature rises we see the X⁻ signal drop significantly at about 55 K which we attribute to electrons escaping their bound trion state due to thermal fluctuations (Supplementary Note 1).

Figure 4b is the zoom-in plot at 15K where we observe slightly different line shapes for X⁻ and X⁰. The X⁰ peak is symmetric showing homogenous thermal broadening effects and is well fit by a hyperbolic secant function which yields a full width half maximum of 5 meV[27,28]. However, the X⁻ peak shows a slightly asymmetric profile with a long low-energy tail consistent with electron recoil effects[27]: The recombination of a X⁻ with momentum $k$ will emit a photon and leave a free electron with the same momentum $k$ due to momentum conservation. From energy conservation, the emitted photon has an energy $\hbar\omega = \hbar\omega_o - \frac{\hbar k^2}{2m_e^{*2}} \frac{M_{X^o}}{M_{X^-}}$, where $\hbar\omega_o$ is the energy of trions with $k = 0$, and $M_{X^o}$ and $M_{X^-}$ are the X⁰ and X⁻ effective masses, respectively. The lineshape of trion PL will thus be the convolution of a symmetric peak function (hyperbolic secant) and an exponential low-energy tail function (Supplementary Figure S5 and Supplementary Note 2). When the temperature is above 70K, we find homogenous broadening dominates over electron recoil and the PL spectrum is fit well by two hyperbolic secant functions (Fig. 4c).

From the fits we extract the X⁻ and X⁰ peak position (Fig. 4d) and the ratio of the integrated intensity of the X⁻ to the X⁰ (Fig. 4e) where we do not present trion data above 150 K because it becomes negligible. We find that the peak positions are fit well (solid line in Fig. 4d) using a standard semiconductor bandgap dependence[29] of $E_g(T) = E_g(0) - S\langle\hbar\omega\rangle\left[\coth\left(\frac{\langle\hbar\omega\rangle}{2kT} - 1\right)\right]$ where $E_g(0)$ is the ground state transition energy at 0 K, $S$ is a dimensionless coupling constant, and $\langle\hbar\omega\rangle$ is an average phonon energy. From the fits we extract for X⁰ (X⁻) the $E_g = 1.657$ (1.625) eV, $S = 1.96$ (2.24) and $\langle\hbar\omega\rangle = 15$ meV for both. Applying our mass action model (Supplementary Figure S3 and Supplementary Note 1) with a trion binding energy of 30 meV results in a good fit to the X⁻:X⁰ intensity ratio (solid line in Fig. 4e).

**Discussion**

In summary, we have shown that monolayer MoSe$_2$ is a true 2D excitonic system which exhibits strong electrostatic tuning of exciton charging via a standard back-gated FET. The observed narrow, well separated spectral features are within the Ti-Sapphire laser spectral range and thus provide remarkable opportunities to selectively probe and control specific excitons using current continuous wave and ultrafast Ti-Sapphire laser technologies. Our results further demonstrate that high quality monolayer dichalcogenides can serve as a platform for investigating excitonic physics and photonic applications in the truly 2D limit with the potential to outperform quasi-2D systems. The results represent unique prospects for this burgeoning class of 2D materials in addition to the recent attention received for their valley physics. Our work expands the horizons for using 2D semiconductors for diverse fundamental studies and technical applications.

Note: During the review process, we became aware of the independent observation of negatively charged exciton in monolayer MoS$_2$[30] and investigation of PL intensity of thin film MoSe$_2$ as a function of temperature[31].

**Methods**

**Sample growth.** MoSe$_2$ single crystals were grown by vapor transport. First, MoSe$_2$ powder was prepared by heating a stoichiometric mixture of Mo (Alfa, 99.999%) powder and Se (Alfa, 99.999%) pieces sealed in a quartz tube under 1/3 atmosphere of pure argon. The ampoule was slowly heated up to 850°C and kept at this temperature for 24 hours. Then, 3.5g of MoSe$_2$ powder was mixed with 0.5g of iodine and sealed in a quartz tube. The sealed ampoule was loaded into a tube furnace for 10 days with the hot zone kept at 1050°C and the growth zone at 1000°C. Plate-like crystals with typical dimensions $6 \times 10 \times 0.1$ mm$^3$ were obtained at the cold end. Room temperature X-ray diffraction confirmed the samples are single phase with 2H structure. Elemental analysis was performed using a Hitachi TM-3000 tabletop electron microscope equipped with a Bruker Quantax 70 energy dispersive x-ray (EDX) system. The analysis confirmed the expected stoichiometric Mo:Se ratio in the crystals.

**Device fabrication.** After mechanical exfoliation of MoSe$_2$ monolayers on SiO2 substrates, a JEOL JBX-6300FS electron beam lithography system was used to pattern contacts in a 300nm bilayer PMMA resist. An e-beam evaporator was used to deposit 6/60 nm of Ti/Au followed by liftoff in hot acetone, cleaning in several IPA baths, and drying under N2.

**Optical measurements.** All spectra were collected via a Princeton Instruments Acton 2500i grating spectrometer and an Andor Newton CCD with samples in a helium flow cryostat. For differential reflectance measurements an Ocean Optics tungsten halogen white light source is reflected into a 40X ultra-long working distance objective using a neutral density (ND), achromatic beam splitter and then focused on the sample (~2 um spot size). The reflected signal from the sample is collected by the objective and sent back through the beam splitter into the spectrometer. For PL measurements, a similar setup is used except the ND beam splitter is replaced by a dichroic beam splitter appropriate for the laser wavelength and a laser line notch filter is inserted before the spectrometer.

**Acknowledgments:**

The authors thank David Cobden and Ming Gong for helpful discussions. This work is mainly supported by the US DoE, BES, Materials Sciences and Engineering Division (DE-SC0008145). HY and WY were supported by Research Grant Council of Hong Kong (HKU706412P). NG, JY, DM, and DX were supported by US DoE, BES, Materials Sciences and Engineering Division. Device fabrication was performed at the University of Washington Microfabrication Facility and NSF-funded Nanotech User Facility.

**Author Contribution**:

X.X. conceived the experiments. J.R. fabricated the devices and performed the measurements, assisted by S.W., A.J. and G.A.. S.W., J.R. and X.X. performed data analysis. H.Y., W.Y. and D.X. contributed to the theoretical explanation. N.G., J.Y. and D.M. synthesized and performed bulk characterization measurements on the MoSe$_2$ crystals. All authors discussed the results and contributed to writing the manuscript.

**Figures:**

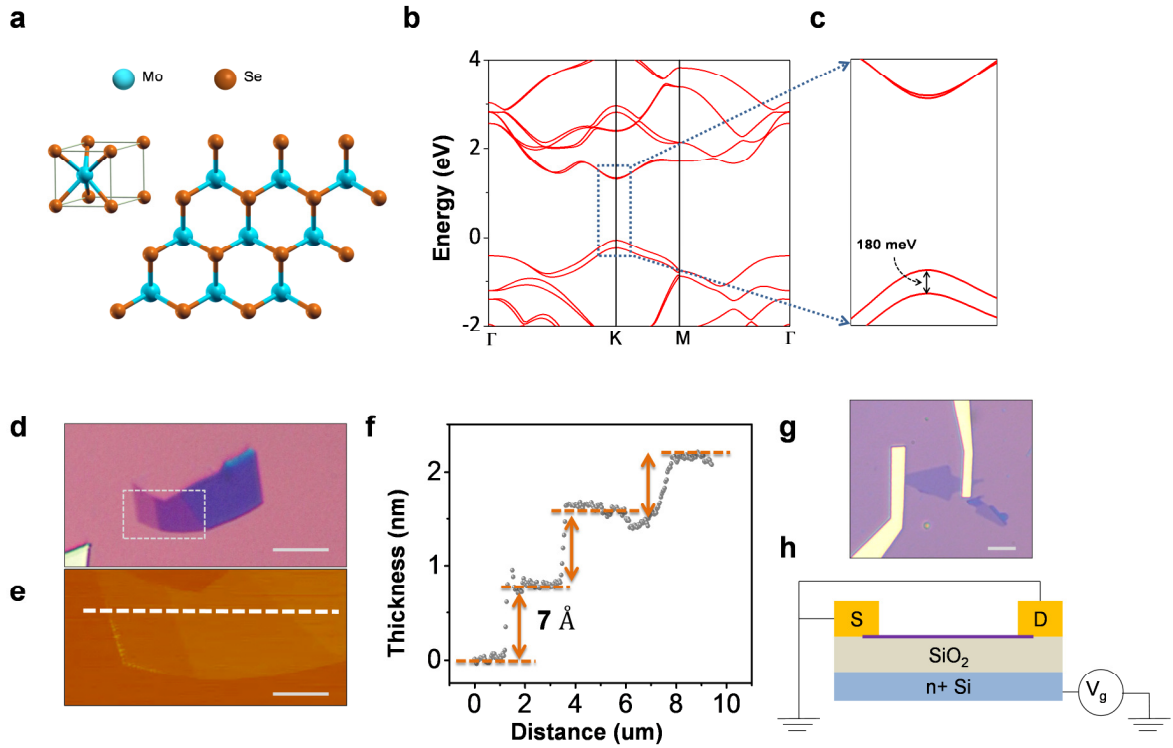

**Figure 1| MoSe$_2$ characteristics and devices. a,** Coordination structure and top view of monolayer MoSe$_2$. **b,** Density function theory calculated band structure. **c,** band structure at K point shows 180 meV valence band splitting due to spin-orbit coupling. **d,** Optical micrograph of exfoliated MoSe$_2$ flake on 300nm SiO$_2$. Scale bar: 5 μm. **e,** atomic force microscope (AFM) image of area highlighted in (d). Scale bar: 1 μm. **f,** AFM line scan along dashed line in (e). **g,** optical micrograph of MoSe$_2$ device. Scale bar: 5 μm. **h,** schematic of back-gated MoSe$_2$ device.

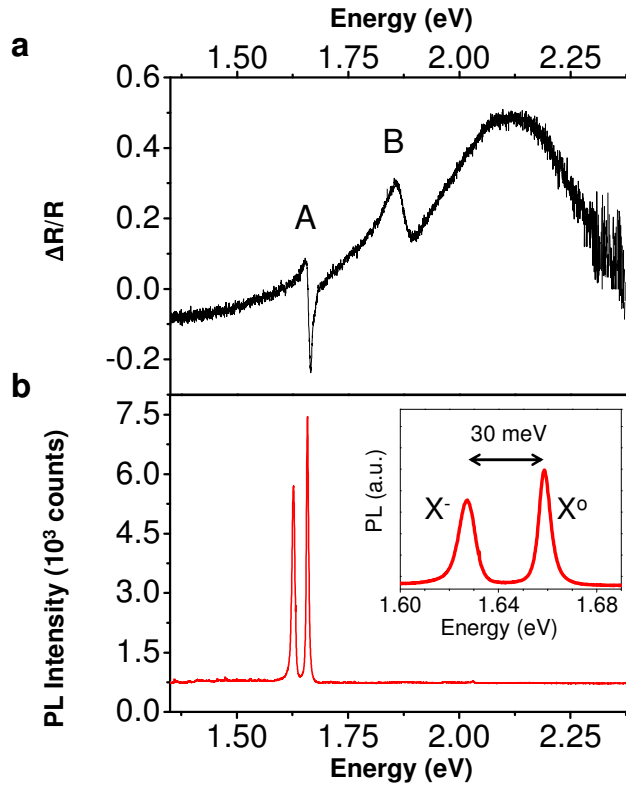

**Figure 2| Differential reflectance and photoluminescence spectra of monolayer $MoSe_2$ at 20 K. a,** Differential reflectance shows A and B excitons. **b,** Photoluminescence (PL) excited by 2.33 eV laser shows neutral exciton ($X^o$) and the lower energy charged exciton ($X^-$). PL from the B exciton has not been observed. Inset: PL of the exciton peaks. The $X^-$ shows a charging energy of about 30 meV.

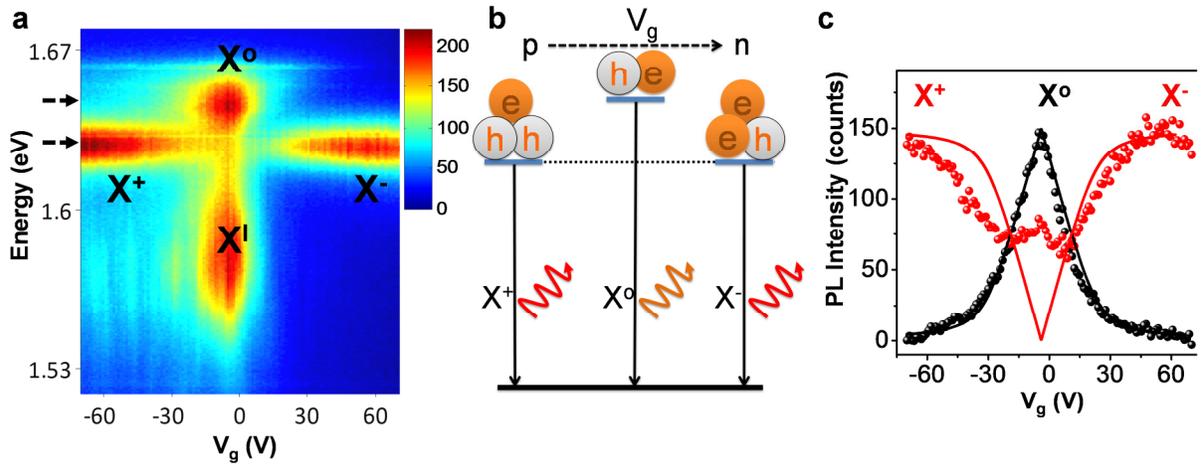

**Figure 3| Electrostatic control of exciton charge. a,** MoSe$_2$ photoluminescence (color scale in counts) is plotted as a function of back-gate voltage. Near zero doping, we observe mostly neutral and impurity-trapped excitons. With large electron (hole) doping, negatively (positively) charged excitons dominate the spectrum. **b,** Illustration of the gate dependent trion and exciton quasi-particles and transitions. **c,** trion and exciton peak intensity vs. gate voltage at dashed arrows in (a). Solid lines are fits based on the mass action model.

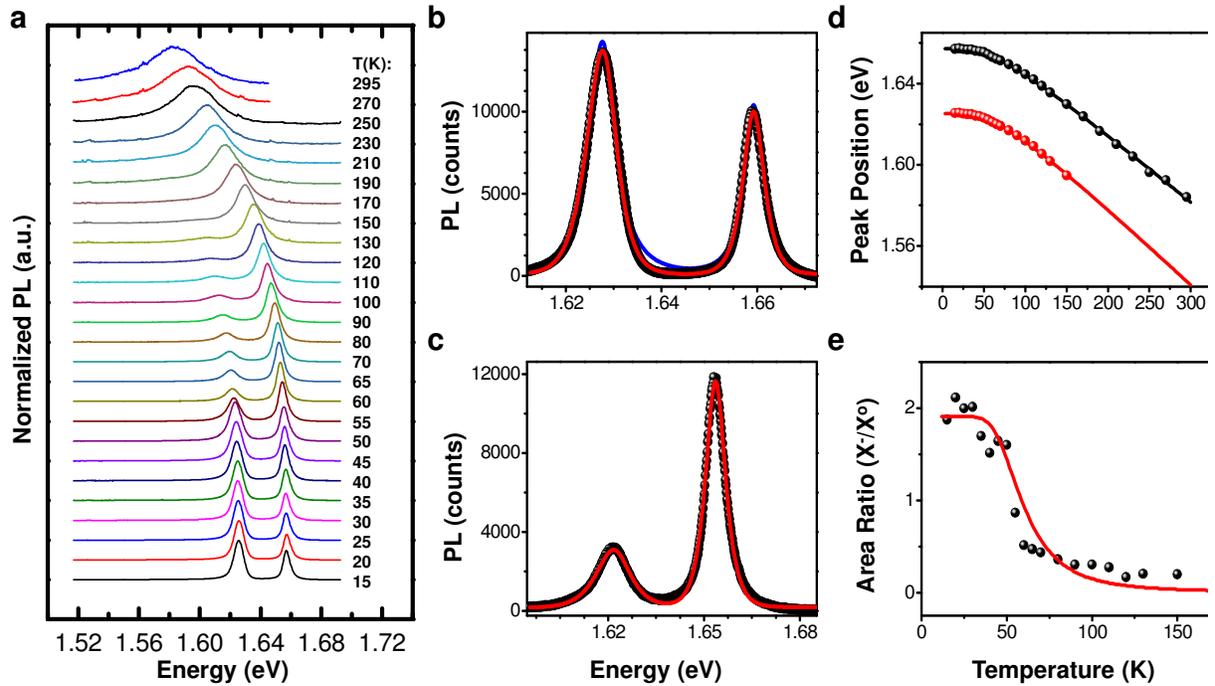

**Figure 4| Temperature dependence of photoluminescence spectrum. a,** Normalized photoluminescence of monolayer MoSe$_2$ vs. temperature. **b,** Line shape fitting at 15 K. Black is data. Red and blue curves are fits with and without considering the electron-recoil effect. **c,** Data and fit at 70 K using two symmetric peaks. **d,** Neutral exciton (black) and trion (red) peak position vs. temperature with fits (solid lines). **e,** Integrated area ratio of trion:exciton vs. temperature with mass action model fitting (red).

# Supplementary Information

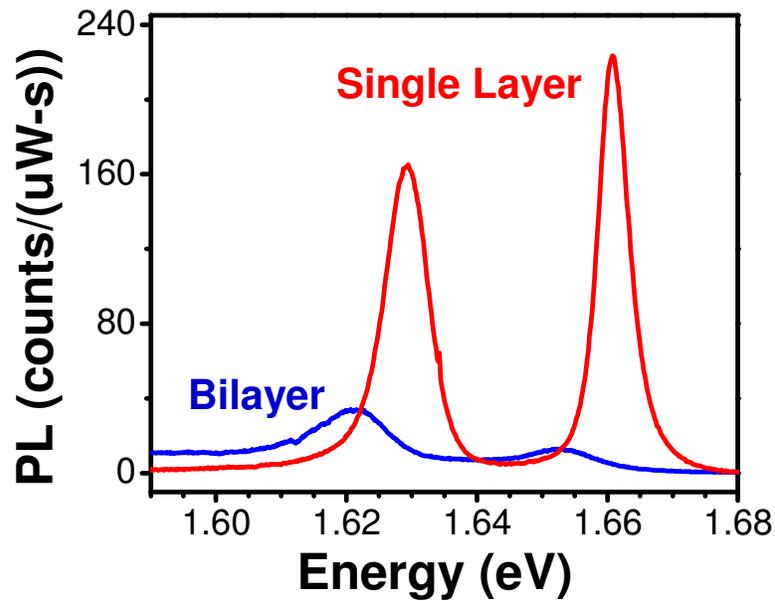

**Supplementary Figure S1: Photoluminescence (PL) of single and bilayer MoSe$_2$.** Both spectrums were taken at 20 K with 532 nm excitation laser. Intensity is normalized to laser power and integration time. Under the same experimental conditions, bilayer samples (confirmed by atomic force microscopy) exhibit suppressed photoluminescence intensity and an overall redshift compared to single layer samples. Like MoS$_2$, these PL features suggest the crossover to direct bandgap at the single layer[10,11] and offer a method of determining the layer thickness.

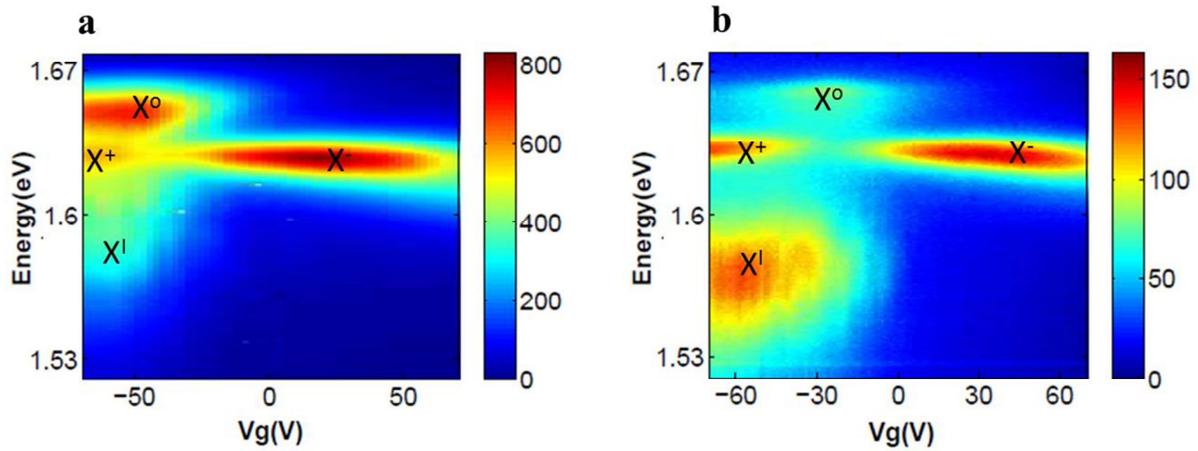

**Supplementary Figure S2: More gate dependence of photoluminescence.** **a,** device D2 under 1.96 eV laser excitation. **b,** device D3 under 1.73 eV laser excitation. Both show nonzero electron doping at back gate $V_g = 0$. In the main text we assume that trions observed in unpatterned samples (no device fabrication after mechanical exfoliation) are negatively charged electron-trions, or $X^-$. These figures provide supporting evidence of this where we see, as with all MoSe$_2$ devices we fabricate, a negative gate voltage is required in order to maximize the $X^o$. Since a maximized $X^o$ corresponds to a minimized carrier density (see Supplemental Figure S3 and Supplementary Note 1), this gate dependence suggests that samples are originally n-doped. This is also consistent with observations[12,14] in MoS$_2$.

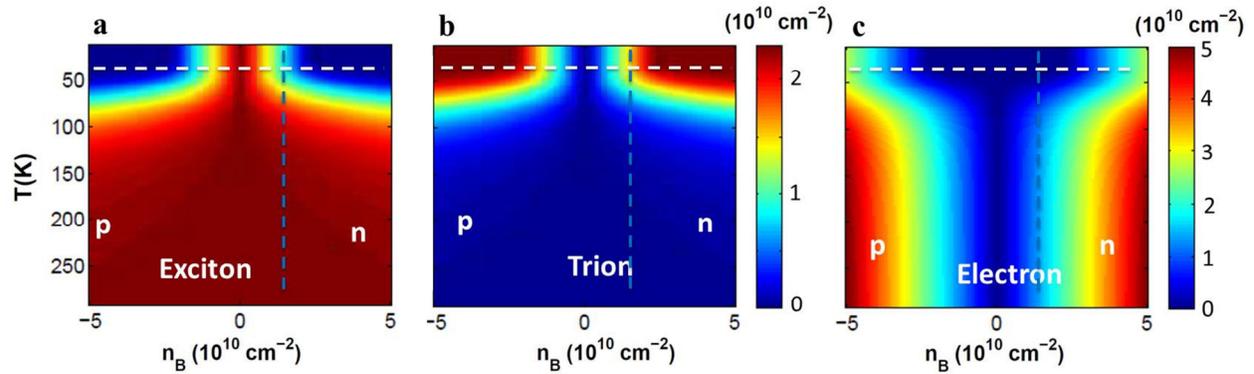

**Supplementary Figure S3: Mass action model plots.** Calculated densities of **a**, exciton, **b**, trion and **c**, electron 2D MoSe$_2$ as a function of background electron density and temperature with a fixed amount of absorbed photons of $3.2 \times 10^{10}$ cm$^{-2}$ during exciton life time. The white and blue lines correspond to the experimental data in the main text for gate and temperature dependence respectively. Details on the model are found in Supplementary Note 1.

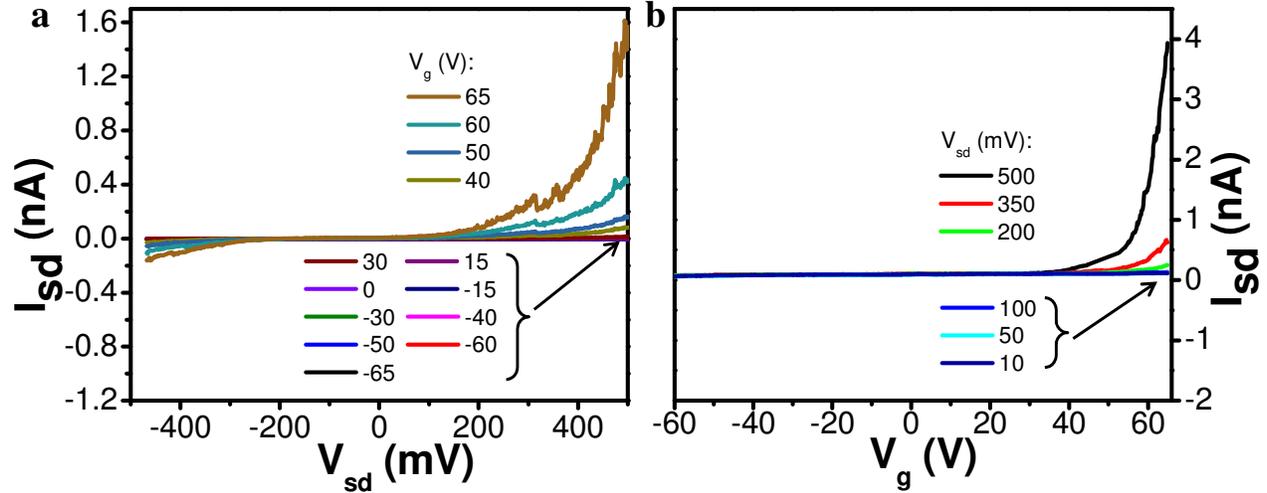

**Supplementary Figure S4: Low temperature transport of single layer MoSe$_2$. a.** Source-drain bias vs. current for various gate voltages. **b.** Gate voltage vs. current for various source-drain biases. The gate dependent photoluminescence (PL) presented in the main text was measured at zero source-drain bias. These figures illustrate that at any gate voltage, there is miniscule conductance for this experimental condition. Specifically, we find the resistance near zero bias to be on the order of 100 GΩ for all gate voltages. Only with appreciable biases above 100 mV do devices show significant conduction. With this in mind, it will be interesting to investigate the gate dependent PL with good metal contacts.

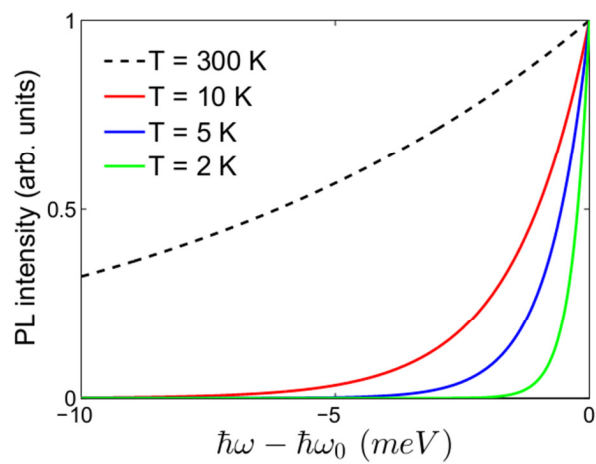

**Supplementary Figure S5: Electron recoil effect on photoluminescence lineshape in trions.** Electron recoil effects result in an exponential lineshape on one side the photoluminescence peaks of trions. Details are described in Supplemental Note 2.

**Supplementary Note 1: Mass action model.**

To determine the relative intensity of the photoluminescence (PL) signals from the various exciton species we adopt a steady state (dynamical equilibrium) model of all the particles in our system. Here, for simplicity, we consider electron-trion and assume that the free hole in the system is negligible. We denote $n_X$, $n_{X^-}$, and $n_e$ for the concentration of $X^o$, $X^-$, and free electrons. $n_P \equiv n_X + n_{X^-}$ denotes the number of photoexcited electrons, and $n_B \equiv n_e + n_{X^-}$ denotes the background electrons (doping level) before light excitation. $n_B$ and $n_P$ are the initial conditions controlled by gate voltage and laser intensity, while $n_X$, $n_{X^-}$, and $n_e$ are steady state variables.

To establish the relationship between these quantities we first write the reaction rate equation for trion formation, $X^o + e^- \rightarrow X^-$. From the law of mass action with trions we have[28,32,33]:

$$\frac{n_X n_e}{n_{X^-}} = A\, k_B T\, \exp(-\frac{E_T}{k_B T})$$

Here $T$ is the temperature, $k_B$ is Boltzmann constant, $E_T$ is the trion binding energy and $A = \frac{4 M_{X^o} m_e}{\pi \hbar^2 M_{X^-}} = 6.18 \times 10^{11} \frac{1}{\text{cm}^2\,\text{meV}}$ in MoSe$_2$ in which $M_{X^o} = m_e + m_h$ and $M_{X^-} = 2m_e + m_h$ are the exciton and trion effective masses respectively. Next, from charge conservation we obtain:

$$n_e + n_X + 2n_{X^-} = n_P + n_B.$$

Solving the above equations gives

$$n_{X^-} = \frac{n_P + n_B + n_A - \sqrt{(n_P + n_B + n_A)^2 - 4 n_P n_B}}{2}$$

Here, $n_A = A\, k_B T\, \exp(-\frac{E_T}{k_B T})$. By plotting these quantities as a function of $T$ and $n_B$, we are able to model the full range of our experimental data (see Supplemental Figure S3). We note that the entire picture and the values we have here for electron-trion can be applied directly to the hole-trion thanks to the same effective mass of electron and hole in our massive Dirac Fermion system. Therefore we have also explained the data in the hole-domain, which is labeled by the negative value of $n_B$ in Supplemental Figure S3.

**Supplementary Note 2: Electron recoil effects.**

Because of energy and momentum conservation, the radiative recombination of an electron- trion with center of mass wave vector **k** results in a conduction electron with wave vector **k** and a photon with energy $\hbar\omega = \hbar\omega_o + \frac{\hbar^2 k^2}{2M_T} - \frac{\hbar^2 k^2}{2m_e^*} = \hbar\omega_o - \frac{\hbar k^2}{2m_e^{*2}}\frac{M_X}{M_T}$, where $M_T = 2m_e^* + m_h^*$ is the trion mass, $M_X = m_e^* + m_h^*$ is the exciton mass, and $\hbar\omega_0$ is the energy of trion with **k** = 0. The photon emission rate is given by the optical matrix element $M(\mathbf{k})$ and trion distribution $f(\mathbf{k})$, which is

$$R(\omega) = \int d\mathbf{k}|M(\mathbf{k})|^2 f(\mathbf{k})\delta(\hbar\omega - \hbar\omega_0 + E(k))$$

where $E(k) \equiv \frac{\hbar k^2}{2m_e^{*2}}\frac{M_X}{M_T}$. In the low density limit, the distribution function $f(\mathbf{k})$ can be approximated by the Boltzmann distribution $f(\mathbf{k}) \propto \exp\left(-\frac{\hbar^2 k^2}{2M_T k_B T}\right)$. The optical matrix element $|M(\mathbf{k})|^2$ has already been numerically evaluated by Stébé et al.[34]. Although for the 2D case they only give results for $\sigma \equiv \frac{m_e^*}{m_h^*} = 0$, 0.1, 0.2, and 0.3, we find those curves can all be well approximated by an exponentially decaying function $|M(\mathbf{k})|^2 = |M(\mathbf{0})|^2 e^{-27.3 E(k)/E_0}$. With $E_0 \equiv 2(1+\sigma)|E_X^{3D}|$, and $E_X^{3D}$ is the 3D exciton binding energy. We expect this equation applies for any $\sigma \in [0,1]$.

The photon emission rate as a function of $\omega$ is then

$$R(\omega) = \int d\mathbf{k}|M(\mathbf{k})|^2 f(\mathbf{k})\delta(\hbar\omega - \hbar\omega_0 + E(k))$$
$$= R(\omega_0) \exp\left[-\left(\frac{27.3}{E_0} + \frac{m_e^*}{M_X}\frac{1}{k_B T}\right)(\hbar\omega_0 - \hbar\omega)\right]\Theta(\omega_0 - \omega)$$

Here $\Theta$ is the Heaviside step function. In the MoSe$_2$ system, $\sigma = 1$ and the 3D exciton binding energy is $E_X^{3D} = R_y^* = 50$ meV. We have plotted the line shape $R(\omega)$ for several temperatures in Supplemental Figure S5. Thus the trion line shape can be fit through convolution of the above contribution with a symmetric broadening effect.

**Supplemental References**